\documentclass[sigconf]{acmart}
\usepackage[T1]{fontenc}
\usepackage[utf8]{inputenc}
\usepackage{babel}
\usepackage[font=small,labelfont=bf,tableposition=top]{caption}
\usepackage{blindtext}
\usepackage{cuted}
\usepackage{float}
\floatstyle{plain}
\restylefloat{figure}
\usepackage{amsmath}

\usepackage{amsthm}
\usepackage{amssymb}
\usepackage{balance}
\usepackage{tabularx}
\usepackage{xcolor}
\usepackage{xspace}
\AtBeginDocument{%
  \providecommand\BibTeX{{%
    \normalfont B\kern-0.5em{\scshape i\kern-0.25em b}\kern-0.8em\TeX}}}

\setcopyright{acmcopyright}
\copyrightyear{2024}
\acmYear{2024}
\acmDOI{XXXXXXX.XXXXXXX}

\acmConference[UIST '24]{ }{October 13--15,
  2024}{Pittsburgh, PA}
\acmPrice{15.00}
\acmISBN{978-1-4503-XXXX-X/18/06}

\newcommand{\sys}{\textsc{Hands-Off}\xspace}

\title{Feminist Interaction Techniques: Deterring Non-Consensual Screenshots with Interaction Techniques}

\begin{document}

\author{Li Qiwei, Francesca Lameiro, Shefali Patel, \\Cristi Isaula-Reyes, Eytan Adar, Eric Gilbert, Sarita Schoenebeck}
\affiliation{%
  \institution{University of Michigan}
  \city{Ann Arbor}
  \country{USA}
}

\thanks{Authors emails: \texttt{rrll,flameiro,shefalip,cisaula,eadar,eegg,yardi@umich.edu} 
\vspace{-0.4cm}
}

\renewcommand{\shortauthors}{Qiwei, et al.}

\begin{strip}
    \centering
    \includegraphics[width=0.65\linewidth, clip, trim=140pt 215pt 120pt 270pt]{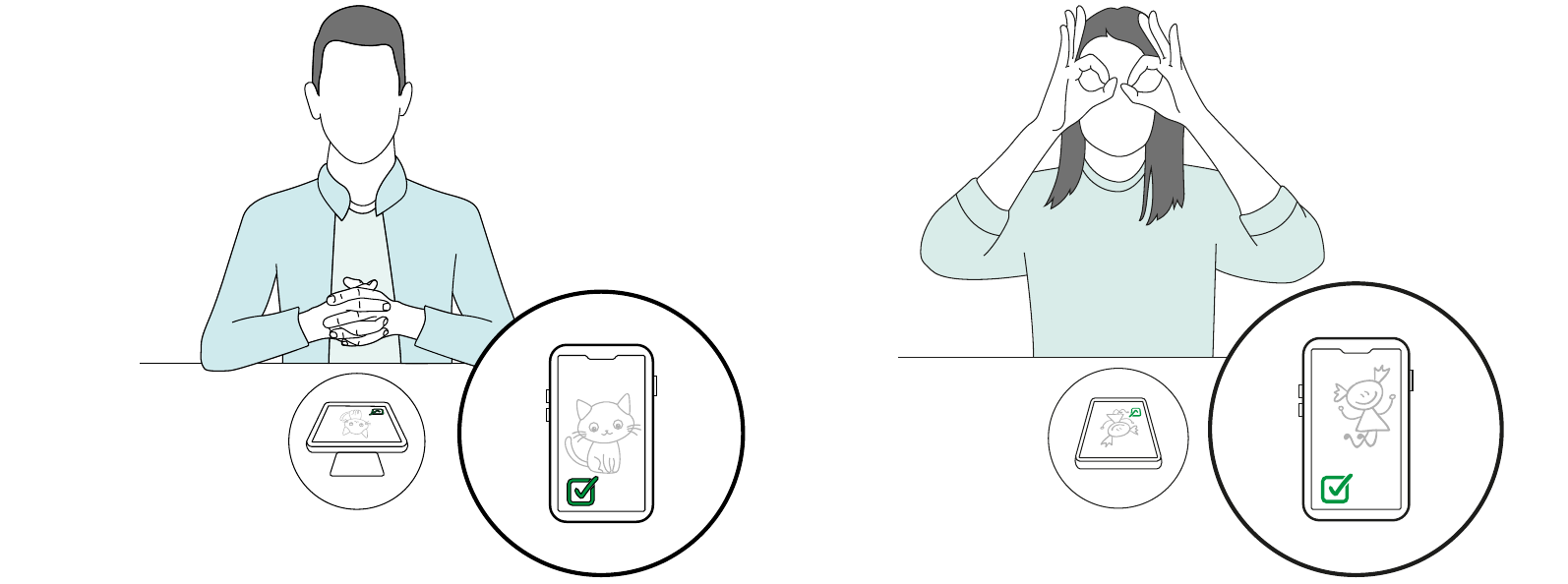}
    \captionof{figure}{Performing interlace and binoculars gestures in \textsc{Hands-Off} to deter non-consensual mobile screenshots.}
    \label{fig:1} 
\end{strip}

\begin{abstract}

Non-consensual Intimate Media (NCIM) refers to the distribution of sexual or intimate content without consent. NCIM is common and causes significant emotional, financial, and reputational harm. We developed \sys, an interaction technique for messaging applications that deters non-consensual screenshots. \sys requires recipients to perform a hand gesture in the air, above the device, to unlock media—which makes simultaneous screenshotting difficult. A lab study shows that \textsc{Hands-Off} gestures are easy to perform and reduce non-consensual screenshots by 67\%. We conclude by generalizing this approach and introduce the idea of \textit{Feminist Interaction Techniques (FIT)}, interaction techniques that encode feminist values and speak to societal problems, and reflect on FIT's opportunities and limitations.

\end{abstract}

\vspace{-5cm}




\maketitle

\vspace{-3mm}
\section{Introduction} \label{introduction}

Can interaction techniques be feminist? Feminist values such as agency, consent, and boundaries seek to create a more just and equitable society. Could low-level interaction techniques--at the boundary between people and machines--embody them? 

Serious societal problems live at this boundary. For example, non-consensual intimate media (NCIM) refers to the distribution of sexual or intimate content without consent. NCIM is prevalent and causes significant emotional, financial, and reputational harm~\cite{henry2020image,citron2014criminalizing,citron2018sexual,franks_drafting_2014,mcglynn_its_2021,qin2024did,walker_systematic_2017,patel_prevalence_2022,ccri2014revenge}. A common NCIM case is when sexual content shared during romantic relationships is duplicated and re-shared without permission.\footnote{This is sometimes referred to as ``revenge porn'', though the term is not appropriate because NCIM does not conform to the definition of pornography.} If an image gets to a recipient's screen, it can always be captured in some way. An image can likely be saved directly, but if it cannot be saved, it can likely be screenshotted~\cite{shore_platform_2023}, and if it cannot be screenshotted, it can be captured using an external camera. 

Offline, learned social rules help govern access to our bodies; in mediated contexts, however, few if any tools exist to express, grant, or revoke consent~\cite{lee_building_2017}. Most existing systems to combat NCIM require reporting the content and then following a set of steps to try to have it taken down, which is reactive and often ineffective \cite{mcglynn_its_2021}. A motivating question for us is: Is it possible to deter non-consensual screenshots with interaction techniques?

To address this issue, and as part of our broader project addressing NCIM, we developed \sys, an interaction technique for messaging applications that deters non-consensual screenshots (see Fig. \ref{fig:1}). After a formative study, we built \sys to require recipients to perform a hand gesture in the air, above the device, to unlock media. Borrowing from the social meaning of gestures (e.g., handshakes)~\cite{muller2013body, zhou2008body}, \sys uses gestures as design friction to convey a user's consent preferences, and acts as a deterrent against screenshots. We built a set of four hand-gesture-based deterrence features. If a sender wishes to share a photo, but only wants the receiver to have controlled, momentary access to it, they can require that the receiver perform a gesture. The media will only reveal itself while the hand-gesture is being performed. We also describe a threat model and design space for how gestures can be used with different content types, relationships, and locations.

Our lab study (\textit{N}=13) shows \sys reduces non-consensual screenshots by 67\% across all gestures---and by 77\% among the top-performing three. The experimental procedure asked participants to view images with \sys, and then try to ``break it'' by obtaining a screenshot of the image. Afterward, we conducted interviews and a usability survey. Participants found the \sys gestures easy to perform, but difficult or ``impossible'' to take a screenshot. Participants resorted to using their feet in some cases (see Fig. \ref{fig:task2}). Critically, participants also reported that \sys establishes social norms about what content the sender wishes to remain transient---setting important social expectations around when duplication is allowed and when it is not. 

Finally, we move toward generalizing the conceptual approach of \sys and introduce the idea of  \textit{Feminist Interaction Techniques (FIT)}---interaction techniques that encode feminist values and speak to societal problems. FIT draws on feminism as a set of generative concepts for computing---such as agency, boundaries, and consent~\cite{bardzell_feminist_2010}. It aims to translate feminism's values---and its goal of a more just and equitable society through them---into lower-level, technical HCI concepts. However, problems like NCIM are complex, deeply-rooted societal issues; technical approaches such as FIT can help, but not ``solve them.'' The oppression of women and gender minorities is not a problem that technology alone can fix. To conclude, we reflect on both FIT's opportunities and limitations.

\vspace{5pt}
\noindent
In summary, the major contributions of this work are:

\begin{itemize}
    \item \textbf{\textit{The \sys interaction technique}} for messaging applications that deter NCIM.
    \item \textbf{\textit{An evaluation of \sys}}. \sys deterred non-consensual screenshots and may create a constructive cycle that establishes social norms around consent.
    \item \textbf{\textit{The concept of Feminist Interaction Techniques (FIT)}}---interaction techniques that encode feminist values and speak to societal problems.
    \item \textbf{\textit{A discussion of the strengths and limitations of FIT}}. NCIM is a deeply entrenched societal problem. Approaches such as FIT can ameliorate NCIM, but not solve it.
\end{itemize}

\section{Related Work} \label{rr}
We review prior work on how values are built into systems. We then summarize prior research systems for visual privacy, placing \sys within a related threat model. Finally, we consider design friction and its potential to shape \sys outcomes.

\subsection{Embedding values in systems}

Values are principles or standards that have some influence over behaviors or decision-making. Within technology, a value refers to the underlying ethics and beliefs that shape the design process and the functions of the technology. These values are not just about aesthetics or usability, but about how the system may impact users and society~\cite{costanza2020design, bardzell_feminist_2010,friedman1996value}. Human-centered computing has sought to design technologies with certain values instilled, and a rich body of work has critically examined values within computing. For example, Costanza-Chock critiqued the implicit gender and body assumptions in modern airport TSA systems~\cite{costanza2020design}, Friedman applied Value Sensitive Design principles for the redesign of the Mozilla browser for cookie awareness~\cite{sengers_reflective_2005, friedman2002informed}, and Fiesler et. al noted feminist values in the redesign of a web page~\cite{fiesler_archive_2016}. 

The concept and practice of feminism has a long and storied history, but has typically prioritized values of empowerment, agency, autonomy, and consent. Bardzell argued in 2010 that the field of feminism had much to contribute to interaction design and called for a deeper engagement with feminist ideas in the field of human-computer interaction~\cite{bardzell_feminist_2010}. That call has been taken up in subsequent work that explores the values embedded in consentful technology, such as control over bodies and affirmative consent in interactions~\cite{lee_building_2017, im_yes_2021}.  

However, building a complex principle like consent into systems is challenging. The ``sociotechnical gap'' describes the gap between the social goals we might have for a system, and what the system is actually able to do in practice~\cite{ackerman2000intellectual}. These challenges are especially felt at the top of the software ``stack'', where human behaviors and interactions create rich, messy social contexts, which computers cannot easily model or build abstractions from. That is, while systems might be pretty effective at collecting and protecting user values at lower layers of the stack, such as operating system level access controls, they struggle with application layer levels, such as content that is shared between people. As Ackerman notes, ``social activity is fluid and nuanced, and this makes systems technically difficult to construct properly''~\cite{ackerman2000intellectual}. Yet, he continues, we can ``augment technical mechanisms with social mechanisms to control, regulate, or encourage behavior.'' This work takes up the charge of augmenting interaction techniques in support of feminist values. %\cite{ackerman2000intellectual,sproull1991connections}. This work seeks to address the sociotechnical gap between 

\subsection{Visual privacy}
 
It has never been easier to take and share photos online. While photo-sharing offers social value to society (e.g. sending a baby photo to grandparents), photo-sharing via wearable devices, ubiquitous surveillance, and facial recognition technologies creates privacy risks. Privacy-inspired systems address this concern by focusing primarily in three areas: physical photography, online photo access, and photo recapture. Prior work includes systems to prevent unwanted photography in public spaces, such as Privacy.tag's QR codes that can be placed on clothing to enact privacy protections~\cite{bo_privacytag_2014} or using smart LED lighting design in physical spaces to prevent photography~\cite{zhu_automating_2017}. The second area of research focuses on limiting users' access to photos. This includes asking individuals depicted in a photo for preferences regarding obfuscation~\cite{olteanu_consensual_2018, shu2017your, shu2018cardea} and drawing cartoons over video streams to protect bystanders' privacy~\cite{hassan_cartooning_2017}, or only showing the faces to friends on social media~\cite{ilia2015face}. A final thread of research includes systems to prevent recapture of a mobile screen, such as preventing second screen capture by changing the lighting and display on a screen~\cite{zhang_kaleido_2015} and digital watermarking systems that collect metadata of the device making screenshots~\cite{gu_anti-screenshot_2023, bai_fast_2023, jiang_information_nodate}, \textcolor{black}{and photos that fade with }time~\cite{pias2022decaying}.

Although screenshots are useful for collecting and sharing information~\cite{cramer_uses_nodate}, their misuse can also have serious consequences. Screenshots play a role in making temporal sharing into a permanent privacy violation~\cite{draper2019corporate, shore_platform_2023}. Illicit duplication and dissemination of screenshots have contributed to mass online harassment, misinformation via decontextualization of content, and revealing personal information that leads to stalking, surveillance, and even kidnapping of both adults and children~\cite{patel_prevalence_2022, brosch2018sharenting, lipton2009we}. While ephemeral social network sites such as Snapchat are often used for temporal and casual communication, they are not ideal for preserving privacy~\cite{pias2022decaying}. Snapchat still allows screenshots and users have discovered additional methods of saving photos and messages sent on the platform~\cite{khan2015snapchat, pias2022decaying}. 

This work is motivated by NCIM, a specific form of privacy leakage harm. The illicit distribution of another individual's intimate content impacts women disproportionately and exemplifies a form of privacy violation that impacts everyone~\cite{citron2014criminalizing, citron2018sexual}. The harms of NCIM are severe and lasting. According to a survey conducted by the Cyber Civil Rights Initiative, of 361 victims of NCIM, 93\% suffered significant emotional distress, 82\% suffered significant impairment in social and occupational areas, 42\% sought out psychological services, and 51\% experienced suicidal ideation \cite{ccri2014revenge}. NCIM can have many causes, including fabrication with deepfakes and recordings during sexual assault. A common occurrence is the non-consensual distribution of intimate content that was initially sent with consent, common in break-ups between romantic couples~\cite{citron2018sexual, franks_drafting_2014, walsh_if_2022, mcglynn_its_2021}. Another form of violation occurs in commercial activities, often involving sex work or the creation of sexual content \cite{qin2024did,mcdonald2021s}, when such content is redistributed or shared without the creator's consent. This can be through copying existing content to a different platform or by sharing privately commissioned work publicly. NCIM is part of a broad class of problems where gender inequalities are perpetuated online, including the creation and dissemination of NCIM~\cite{citron2014criminalizing,citron2018sexual, citron2022fight, franks_drafting_2014, geeng2020usable, powell_image-based_2019, mcglynn_its_2021, freed_digital_2017, qin2024did},  our data bodies~\cite{lee_building_2017}, acts of intimate partner violence~\cite{tseng2020tools, zou2021role, freed2018stalker, freed_digital_2017}, and other forms of online harms. These problems go far beyond privacy breaches, causing emotional, reputational, and economic harm.

\subsection{Design friction}

HCI has historically focused on making systems that are intuitive, unburdensome, and that prioritize performance for the end-user. Introducing friction at the cost of lessening performance may increase desired behaviors, similarly to how a stop sign slows driving but increases safety. Design friction is the concept of deliberately making some interactions more difficult, in order to make them more purposeful, safe, and thoughtful~\cite{gould_special_2021}. 

Design friction has been implemented in social media platforms to avoid unsafe or undesired actions, particularly ones that have downstream harms on other users. Wang et. al deployed a countdown timer on Facebook to ensure the user waits 20 seconds before making a post in order to minimize regrettable posts~\cite{wang_field_2014}. Gould et. al notes several design friction choices on Twitter's platform to warn users against potential misinformation in some posts, as well as asking users to confirm they have read an article themselves before re-tweeting~\cite{gould_special_2021}. 

Friction can also be a tool towards meaningful reflection and intention. It is common for functionally opposing design approaches to remove key aspects of their function in order to highlight their values. This includes works in critical design, counter-functional design, slow design, reflective design, and lucid design ~\cite{pierce_counterfunctional_2014, sengers_reflective_2005, gaver2000alternatives}. Physical friction can also be designed to induce user reflection. For example, ``Crank That" is a physical hand-crank intervention to increase friction and awareness of passive consumption in Twitter content feeds~\cite{song_crank_2021}. Users are required to continuously turn a hand crank in order to power a screen to use social media, allowing users to ``question how they value various interactions on social media by transforming their usage into an embodied one in which they have to physically work for their experience''~\cite{song_crank_2021}. 

Our work applies design friction to encourage users to think about consent when taking screenshots in online conversations. We leverage intentional design barriers to promote responsible use. We also shift the burden from those who are the victims of consent violations towards those who might be the perpetrators of consent violations. That is, instead of responding to non-consensual sharing reactively (e.g. requesting a takedown), we intervene in it proactively (e.g. deterring from violations). 

An effective proactive mechanism should prevent or deter unwanted behavior. Currently, only a few messaging platforms allow users to prevent screenshots altogether.\footnote{https://www.safedigitalintimacy.org/state-of-the-industry} StopNCII, a hashing-based system developed by multiple stakeholders including Meta, allows users to upload photos to a server before these photos are ever posted by someone else, and then block subsequent attempts by someone else to upload these photos.\footnote{https://stopncii.org/} However, even then there are many drawbacks. Transformed images may evade the hash algorithm. When these images continue to be reposted, platforms may overly rely on StopNCII and invest less effort in removals. And of course, users must trust the third party--in this case, Meta. Our work aims to make the burden more equitable, placing ``the work'' on a potential perpetrator rather than the potential victim.

\section{Formative Study} \label{formative}
We conducted a formative interview study to develop design requirements for a technological intervention. The formative study aimed to augment prior qualitative and survey research on NCIM~\cite{citron2014criminalizing,citron2018sexual, citron2022fight, franks_drafting_2014, geeng2020usable, powell_image-based_2019, mcglynn_its_2021, freed_digital_2017, qin2024did}. The first portion of the interview focused on participants' existing screenshot experiences and practices. The second portion asked participants for their thoughts about using a screenshot deterrence system. In our conception of \sys, we believed that a gestural interface--one that prevented viewers from using the screenshot features by occupying their hands--could be an effective deterrent. We used the formative study as an opportunity to test this idea. Thus, in the second part of the interview, we had participants explore how they would use a screenshot deterrence feature. We described the design of a gesture-based approach and solicited reflection and feedback feedback. Our university's IRB exempted the study.

\subsection{Participants}

Eight participants were recruited from two undergrad courses at a large university, \textcolor{black}{consisting of 4 women and 4 men}. We specifically recruited participants who were able and willing to share their own or other peoples' experiences about screenshots. \textcolor{black}{We wanted to capture not only NCIM-specific experiences but also wider patterns of unauthorized screenshot sharing so that Hands-Off may be a system used for messaging broadly. We gave participants the option of either 1:1 solo interviews or paired interviews in which the participant brought a friend, who often acted as a confidante, to encourage more conversation.} Each interview session took between 45 to 60 minutes. \textcolor{black}{Each participant was compensated \$15 for a pair interview or \$20 for a solo interview. In the paired interview, both participants were paid and both generated information used for the study. }

\subsection{Interview results}

Two authors analyzed the interview data. We annotated interview transcripts and used open coding to find themes across all sessions. Below are high-level themes from our formative study.

\subsubsection{Theme 1: The need for screenshot prevention}

Participants wanted enhanced privacy measures in mobile messaging platforms, especially in scenarios where sensitive content is exchanged. The dissemination of sensitive content without the sender's knowledge can have far-reaching consequences. A participant shared an incident that exposed a critical gap in notification systems, leaving the victim unaware of unauthorized sharing.

\begin{quote}
    \textit{``A girl was sexting to a friend of mine. And one of my friends sends those pictures to, like, half of the group or whatever. And then I think someone inside the group sent it to another person. It went through most of our grades. I don't know if the girl knew who did that. I mean, there was like, no notification. So she just came into the situation where those photos were running through the class.'' –P4} 
\end{quote}

Furthermore, harm can arise from uncontrolled text message re-sharing. Participant 2 recounted a personal experience of sharing a screenshot of a conversation with a professor, only to have it forwarded to the entire class. 

\begin{quote}
    \textit{``I had shared my conversation with a professor with someone, and I didn't really want everyone to know about it. But that person forwarded it to the entire class. And after I found out ... I was angry. The effects of that incident continued till the day I graduated. People in that group forwarded it to their friends and spoke about me. Not just my department, but other people also would come up to me and ask me about it.'' —P2}
\end{quote}

\subsubsection{Theme 2: Perceived security of gesture-based screenshot prevention}

We presented participants with a possible gesture-based approach, using hand gestures to both indicate preferences about and deter unwanted screenshots. Participants expressed a strong preference for gesture-based security features within mobile messaging applications. Their feedback shed light on the underlying motivations and practical benefits. Gesture-based security provides a tangible barrier, making it significantly harder for the recipient to share content.

One participant conveyed the importance of a physical safeguard that actively prevents sharing, especially in scenarios involving acquaintances and friends. 

\begin{quote}
    \textit{``Even if there was just a feature that said, this person doesn't want to share with an acquaintance, I wouldn't necessarily trust them to actually do that. And so I would want them to do something that physically prevented them from doing that.'' —P4} 
\end{quote}

Participants noted that gesture-based deterrence could reduce the potential awkwardness associated with directly requesting confidentiality through written or spoken messages. This suggests that a gestural approach may enhance privacy while streamlining communication.

\subsubsection{Theme 3: Gesture should depend on context}

All participants believed that there should be a variety of gestures to choose from to account for various usage contexts. Specifically, context depends on three axes: the type of content sent, the relationship with the recipient, and the recipient's location of viewing. Table \ref{tab:design_axels} maps the connection between the axis and its design considerations.

\begin{table}
    \centering
    \begin{tabular}{p{0.25\linewidth}|p{0.65\linewidth}} 
         \textbf{Axis} & \textbf{Design Considerations} \\ \hline 
         Content & Serious content\textrightarrow serious gesture \newline Silly content\textrightarrow silly gesture\\ \hline 
         Relationship & Close\textrightarrow simple, less protective gesture \newline Not close\textrightarrow more protective gesture\\ \hline 
         Location & Public\textrightarrow socially acceptable gesture \newline Private\textrightarrow any gesture 
    \end{tabular}
    \vspace{3pt}
    \caption{Design considerations from formative study. The type of gesture matters for of content sent, the relationship between sender and recipient, and the location for viewing the content.}
    \label{tab:design_axels}
\end{table}

\textbf{Type of content sent.} Participants emphasized the importance of tailoring gestures to the nature of the content being shared. For instance, for lighthearted or silly content, a corresponding playful or casual gesture was suggested. In contrast, when sharing serious or sensitive information, participants favored a more solemn and meaningful gesture that aligned with the gravity of the content:
\begin{quote}
     \textit{``What you're doing with your body elicits of obviously different emotions \dots so it would be really hard for me to engage with a serious text effectively if I had to do something silly.'' –P5} 
\end{quote}

\textbf{Relationship with recipient.} The relationship with the recipient mattered in determining the appropriateness and complexity of the gesture. For close relationships, a simpler, less protective gesture was suitable. Conversely, for interactions with acquaintances or less familiar contacts, participants leaned towards employing a more intricate and potentially challenging gesture to enhance security.

\textbf{Location of viewing.}
Participants thought the viewing environment was a crucial factor in gesture design. In public spaces, there was a preference for gestures that blended seamlessly with everyday actions. This choice was motivated by the desire for subtlety and to avoid drawing attention. In private settings, however, participants felt more at liberty to employ a wider range of gestures. 

\section{System Design} \label{system}

\begin{figure}
    \centering
    \includegraphics[width = 0.95\linewidth, clip, trim= 280pt 50pt 150pt 150pt]{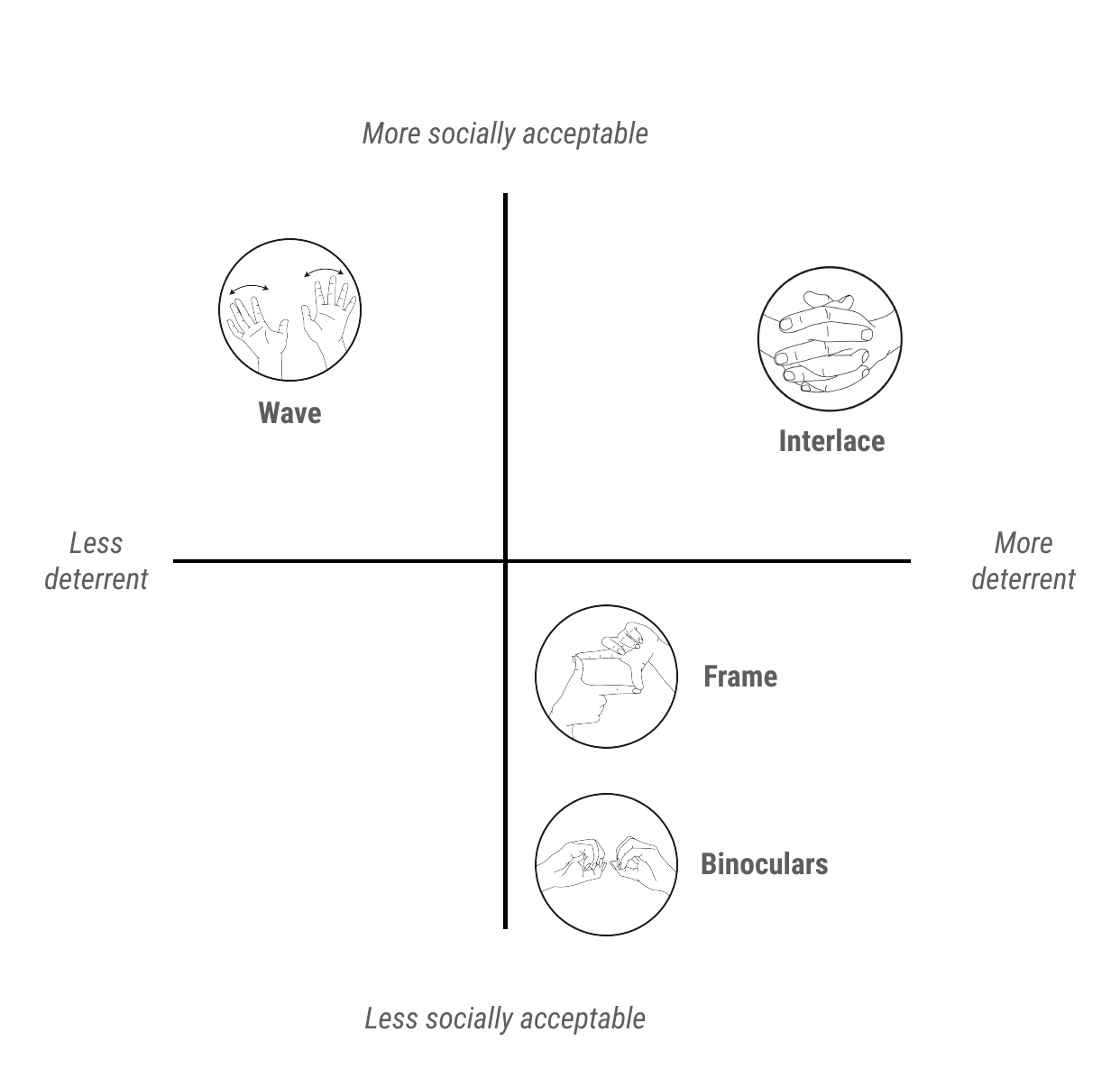}
    \captionof{figure}{A priori, designer-view of deterrence vs social acceptability for \textsc{Hands-Off} gestures.}
    \label{fig:designspace}
\end{figure}

\sys gestures are designed around principles of deterrence and social acceptability, with each usage context leading to the design requirements of a particular gesture. We describe a set of design goals (see Fig. \ref{fig:designspace}) and threat model (see Fig. \ref{fig:threatmodel}) for \sys gestures. We then illustrate how this design space maps to specific use cases and conclude with implementation details for \sys.

\subsection{Threats and design goals}

\subsubsection{Threat model}

\begin{figure*}[t]
    \centering
    \includegraphics[width=0.80\linewidth, clip, trim= 30pt 300pt 40pt 120pt]{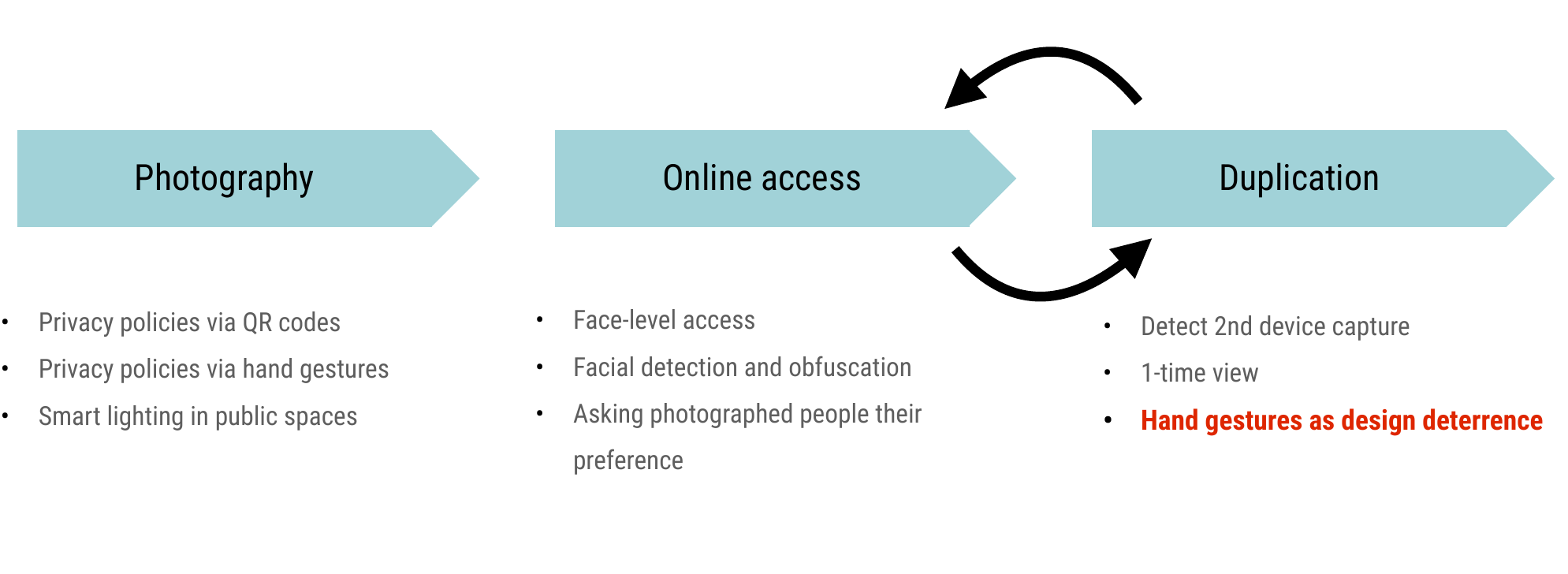}
        \captionof{figure}{Threat model of possible methods of screen duplication after content is sent. Duplicated content may be posted online repeatedly, continuing this cycle. \sys contributes to deterrence against screenshotting content. \textcolor{black}{Also listed is prior work and where it fits in this threat model: preventing photography(\cite{bo_privacytag_2014, zhu_automating_2017}), preventing online access(\cite{ilia_faceoff_2015, liu_hideseeker_2022, hassan_cartooning_2017, vishwamitra2022towards}), and preventing duplication(\cite{zhang_kaleido_2015, pias2022decaying}).}}
    \label{fig:threatmodel}
\end{figure*}

The threat model for \sys is unwanted photo duplication in online sharing~\cite{bedi2013threat} (see  Fig. \ref{fig:threatmodel}). The system should deter general unauthorized duplication, which would take media out of its original context. We borrow from prior visual privacy systems and ``assume that the adversaries are the users, the online services and third parties (e.g., external observers)''~\cite{olteanu_consensual_2018}. The paradigm begins with a sender who sends a piece of original content to a recipient. From there, the recipient may download the file, screen record, screenshot, capture it using a second device themselves, or utilize an accomplice to capture on a different device.

\sys goals include \textit{preventing} screenshots, and \textit{deterring} second device capture by the same individual. Some portions of the threat model are accounted for in prior work. In-app downloads and screen recording may be disabled by the system developer, and second device capture may be deterred with a system such as Kaleido that uses lighting patterns to disrupt outside cameras~\cite{zhang_kaleido_2015}. Screenshots on the device may also be prevented within the operating system of mobile devices, but this is rarely implemented in popular mobile messaging applications. As we have observed, there is no approach that can absolutely \textit{guarantee} that an image not be captured and reshared. However, it is worth considering which approaches raise the cost of different types of ``attacks''.

\subsubsection{Design goals} \label{designgoals}

\sys functions by occupying both hands to prevent individuals from engaging the screenshot features on the device. As the hands are occupied, this may further deter second device capture.\footnote{Clearly, this is just a deterrent. A particularly motivated attacker could position a second device, record a video, clip frames, and recover a fairly good representation.} However, our analysis of themes in the formative study reveals additional specific design constraints: 

\begin{itemize}
    \item \textbf{D1:} Gestures should be easy to perform and detectable.
    \item \textbf{D2:} Gestures should, at minimum, deter screenshot capture.
    \item \textbf{D3:} Gestures should suit different viewing contexts, content types, and receiver-sender relationships.
\end{itemize}

Because our focus is on using the hands as \textit{deterrence} rather than the traditional \textit{input}, our design space for hand gestures is distinct from those defined in prior work~\cite{wobbrock_user-defined_2009,wobbrock_gestures_2007}. For this reason, we chose to create our own inclusion criteria. The design guidelines allow us to focus on specific gesture types while eliminating others. For example, an alternative approach is to use multi-touch detection on the screen itself. This can engage the hands in a way that prevents screenshots. However, these may be harder to execute and detect, requiring contortion of the hands on the device (violation of D1); may be less secure as the hands are close to the buttons (violation of D2); and because the hands are on the device, they may function \textit{too well} across all contexts (violation of D3). Dynamic gestures (e.g., waving, clapping, etc.) may be easy to perform but harder to detect and may limit the contexts in which they are applied (e.g., are less suitable in a public setting).

In addition to the explicit guidelines above, we have opted to focus on abstract gestures with relatively little cultural significance. Some gestures are culturally dependent in meaning (e.g., palms touching in a prayer gesture) whereas others have existing iconic or metaphoric roles (e.g., hands making a heart symbol). By relying on more abstract gestures, we hope that their meaning may be more flexible. This allows some degree of communicable understanding through gestures, while still allowing flexibility for how users choose to interpret them. That said, depending on context, gestures with existing meaning may be desirable for certain applications. These are possible with \sys, but are left to future work.

\begin{figure}
    \centering
    \includegraphics[width = 0.85\linewidth]{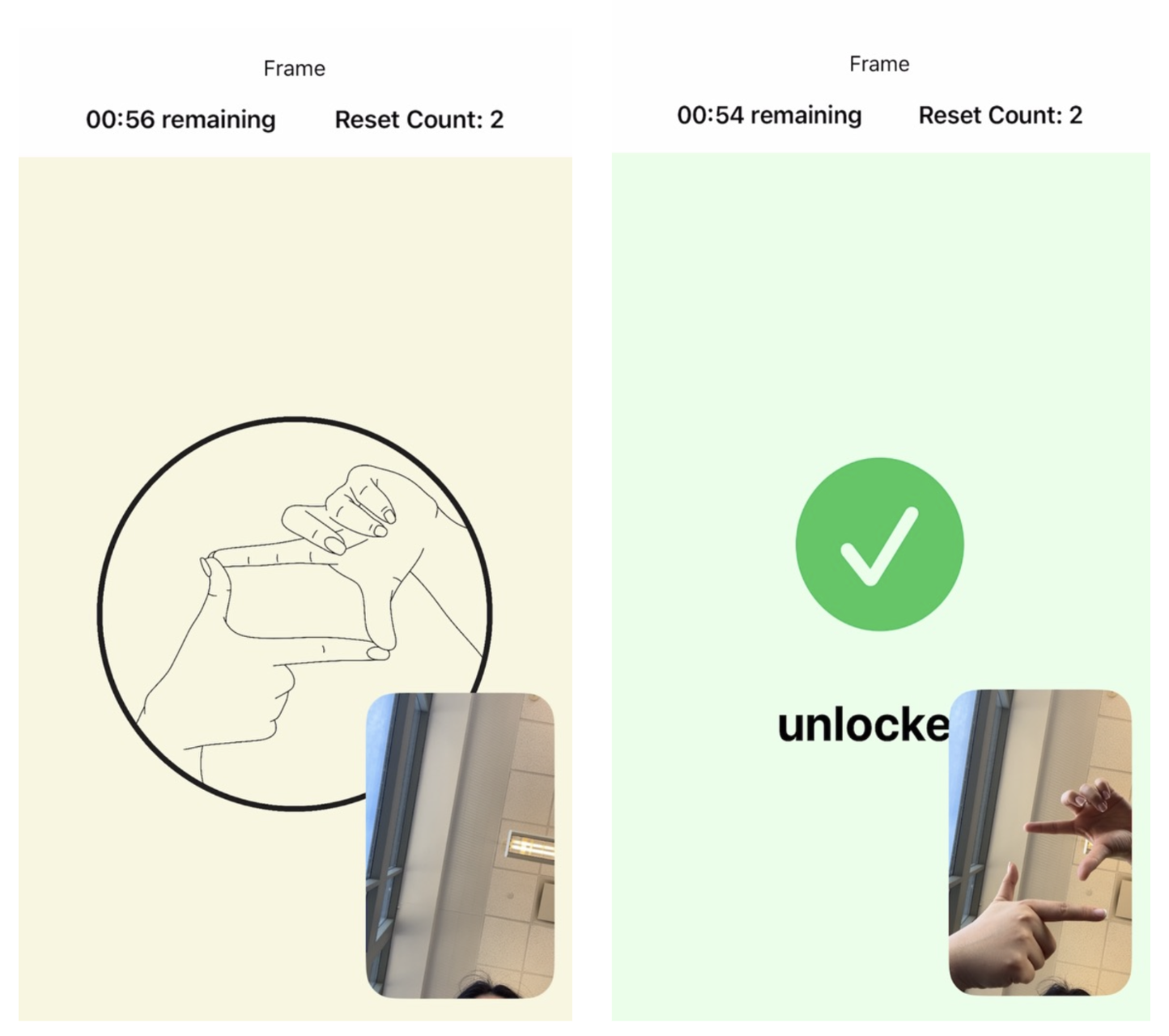}
    \captionof{figure}{\sys interface we use in the usability study. The left screen shows users what gesture to perform, and the right screen shows a successfully unlocked screen after the gesture is detected by the model.}
    \label{fig:system 1}
\end{figure}

\begin{figure}
    \centering
    \includegraphics[width=\linewidth, clip, trim= 15pt 30pt 30pt 15pt]{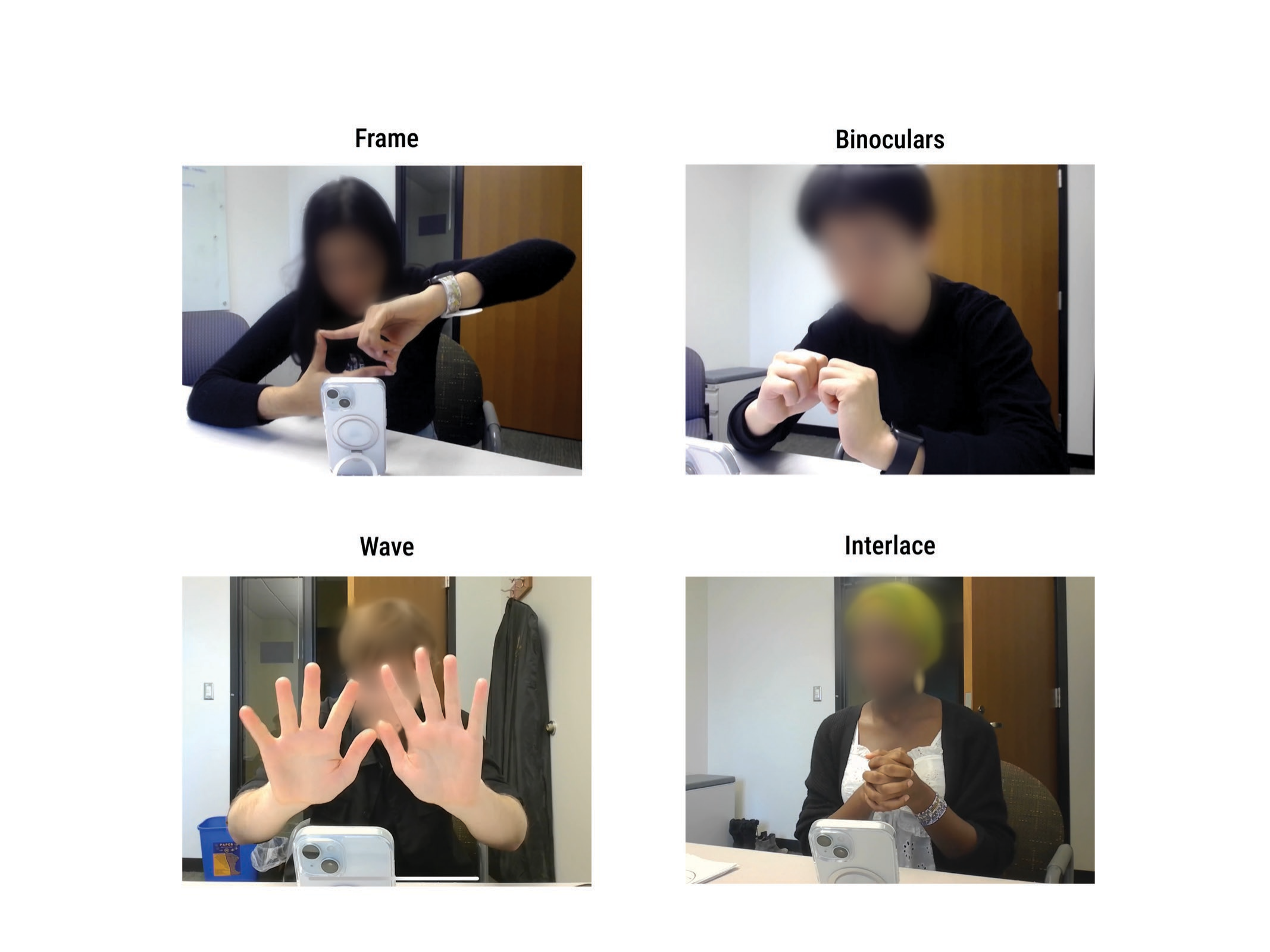}
    \captionof{figure}{Participants performing the \sys gestures.}
    \label{fig:task1}
\end{figure}

\subsection{Design space}

\sys gestures are shaped with two key factors in mind: preventing misuse and fitting within societal norms. Like a literary symbol, social gestures are imbued with meaning. They communicate through physicalization of language or ceremony~\cite{kang2016hands, schiffrin1974handwork}. A wave goodbye between friends, a handshake after a negotiation, or a raised hand in a court of law are all means of communicating between parties but also to observers. They have semantic meaning~\cite{xu2009symbolic}, which may be learned, innate~\cite{iverson1997s}, or inferred (e.g., hands with palms outward to represent a `wall')~\cite{muller2013body}. Even a gesture made in private can embody a sentiment or belief or serve a regulatory function~\cite{rodriguez2007private}. Using a gesture, even an abstract one, can physicalize an affirmation, whether observed by someone else or not: ``I understand that the image was shared with me and is not intended for sharing with others.'' The friction imposed by the physical nature of the gesture further acts to reinforce the meaning and, thus, a broader value system of consent, autonomy, and agency. It is not a button to be clicked away.

As a form of body language, the choice of gestures in interaction must take into account likely social interpretation. This is key as \sys exists within two overlapping, but distinct, design spaces: bodily friction and social friction. Bodily friction is the invisible but physical restraint. The constrained space is the gap between the user when performing the necessary gesture, and the physical location of the two phone buttons that need to be pressed simultaneously to screenshot. In \sys, the gap is maintained by the system's detection of a hand gesture and the distance between reaching over and pressing the two buttons. Along the social dimension, someone using their phone in a public space will consider their social appearance in doing so. Prior work has shown that people generally want to avoid making social blunders in the presence of others, and subtle movements and movements that look and feel like everyday actions are preferred over uncommon movements~\cite{rico2010usable}.

\subsubsection{\sys gestures.} \sys's gestures map to different locations along hypothetical deterrence and social acceptability axes (see Fig. \ref{fig:designspace}). \textit{Wave:} Waving two hands at a phone in public is common when we are greeting someone on video chat, and thus blends in as socially acceptable. \textit{Interlace:} Two hands each holding the other with fingers interwoven for additional security, and is also a `low-key' gesture to do in public settings. \textit{Frame:} Viewing something through frames that the fingers create lends an element of playfulness, and is somewhat silly to do in public. \textit{Binoculars:} This is the strangest and silliest of the four hand gestures, and is likely preferred to be done in private only.

\subsection{Use cases}

\sys may be used in any scenario when someone may choose to share a message or photo, and does not want the recipient to have permanent access to it. Our interview study shows that users may choose to use \sys to share content such as personal documents, emotionally intimate moments, playful memories, or romantic photos. 

%Figure \ref{fig:mockup} shows a mock-up with \sys incorporated within a fictional application myMessage. 

\textbf{Silly moment just for us.} Sam is having a good time and wants to share a photo of him out drinking with his long-distance best friend Ray. However, Sam doesn't want Ray to send this photo to anyone else. He chooses \sys and selects the \textit{binoculars} option to match the silliness of the photo with the gesture, and also to imply Ray should not reshare this photo. When Ray receives the message, he is prompted to perform the hand gesture and then sees the photo Sam has shared. 

\textbf{Romantic intimacy.} Anna met Jordan on a dating app recently and they have been chatting online. They begin sexting and Anna wants to send some photos. However, she doesn't know Jordan well and wants to take protective measures before doing so. She only wants Jordan to see these photos, not anyone in Jordan's group chat. She chooses the \textit{frame} option to send a photo to Jordan. 

\textbf{Office gossip.} Ali and Jo are best friends at work. Today, Jo heard some very interesting gossip that she wants to share with Ali. It's somewhat sensitive and she wants to ensure he can't pass on the information -- at least not with proof that she had shared it. Jo selects \textit{interlace} because it matches the nature of the topic. 

\begin{table}
    \centering
    \begin{tabular}{p{0.15\linewidth}|>{\raggedleft\arraybackslash}p{0.14\linewidth}|>{\raggedleft\arraybackslash}p{0.15\linewidth}|>{\raggedleft\arraybackslash}p{0.15\linewidth}|>{\raggedleft\arraybackslash}p{0.11\linewidth}|>{\raggedleft\arraybackslash}p{0.05\linewidth}} 
         \textbf{Gesture} & \textbf{False Positives} & \textbf{False Negatives} & \textbf{Precision} & \textbf{Recall} & \textbf{F$_1$}
         \\ \hline 
         Wave & 3 & 1 & 98\% & 99\% & 0.98 \\ \hline 
         Frame & 6 & 9 & 97\% & 95\% & 0.96 \\ \hline 
         Interlace & 32 & 2 & 86\% & 99\% & 0.92 \\ \hline 
         Binoculars & 4 & 6 & 97\% & 96\% & 0.96
    \end{tabular}
    \vspace{3pt}
    \caption{Model shows high precision across 3 of the 4 gestures. Interlace was an exception because the model had difficulty recognizing interlaced fingers as two hands instead of one.}
    \label{tab:ml}
\end{table}

\subsection{Implementation details}
%We do not build a full message sharing application but instead focus on the gesture detection and screenshot deterrence to isolate the novel aspects of this work.

%% optimizations, tradeoffs, rationales for choices, engineering process 
We built \sys using SwiftUI for iOS devices. We chose the iPhone for its consistent button placement and hardware design. We rely on pose detection using a custom CoreML hand pose detection model trained with a custom corpus we built. We defined a multi-class neural-network classification model with 20 convolution layers for interlace, wave, frame, binoculars, and a background class for when none of the above gestures are detected. Training data consisted of 120 to 200 photos of hands performing each gesture per class. We collected the vast majority of the data ourselves by taking photos of the hands of participants. We account for high and low lighting, light and dark skin tones, and different photography angles. The remaining photos are copyright-free images of hands from the internet. The model was trained to recognize the hands through the positions of the fingertips and the locations of the joints. See Table \ref{tab:ml} for model performance. Facial data was not used to train the model.

Gesture detection in the \sys system initially shows a screen with the hand gesture that the user is prompted to perform as in Figure \ref{fig:system 1}. The system can detect hands while the phone is on a flat surface or propped up (e.g., with a phone stand). Once the model detects the hand gesture successfully, the underlying photo is revealed. The camera feed is continuous, and the photo is covered again as soon as the hand gesture stops being detected. To decrease the times that the image flickers between covered and uncovered, we add an adjustable confidence level for the gesture detection model's predictions with a default set at 90\%. The rest of the user interface is composed of a front-camera view at the bottom right corner similar to those found in common video chat software. 

\begin{figure*}[t]
    \centering
    \includegraphics[width=0.9\linewidth]{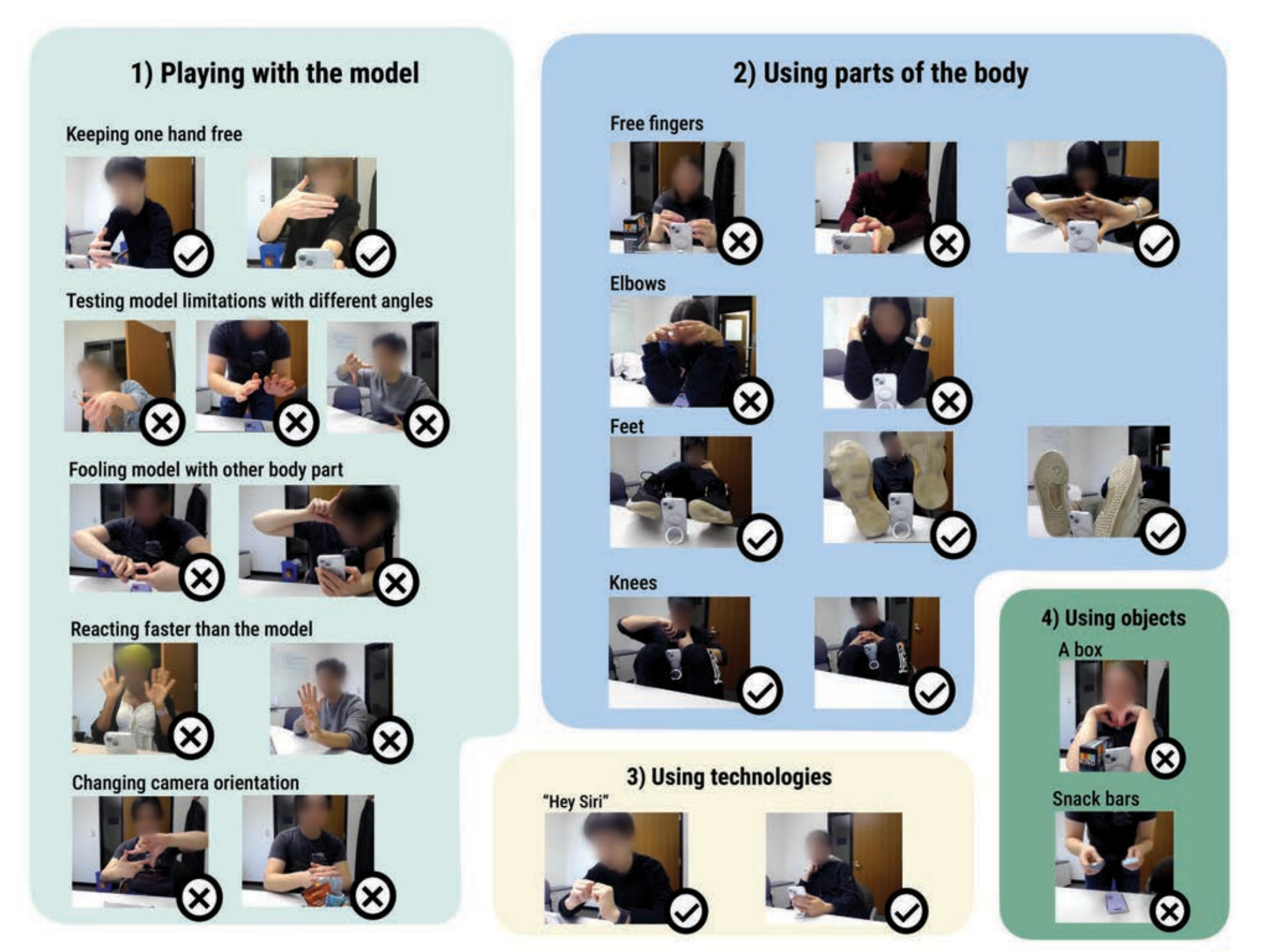} %
    \captionof{figure}{Attempts to break \sys surfaced many creative methods, including playing with model boundaries, using body parts, voice-based technologies, and prop objects.}
    \label{fig:task2} 
\end{figure*}

\subsection{\sys limitations}

\subsubsection{System limitations.}
One limitation of using the CoreML model is it consistently recognized interlaced hands as one hand instead of two, which impacted precision for interlace (see Table \ref{tab:ml}). For each other hand gesture, \sys required two hands detected as well as gesture detected to display the photo. However, we had to silently change this two-hands criterion for the interlace option. This limitation tells us the CoreML vision model works well for detected separated hands or poses, but may be limited for gestures when the hands stack or otherwise overlay each other. 

We note several limitations in \sys. First, \sys is based on assumptions and modes of interactions specific to the iPhone. Screenshot deterrence is built on top of one static assumption of the double button press on the left and right sides of the phone body. The methods of screenshots can vary in other designs of mobile phones. As future work progresses, limitations may also arise in securing hand gesture-based consent in AR, VR, and XR spaces due to the possibility of certain gestures already being used as input by the specific system. Second, this study tackles just one aspect of many potential privacy concerns in the threat model. Screen recording can still be used to capture the screen if that feature is not disabled in the application. Future research in gesture-based deterrence should consider allowing users to create their own custom hand gestures. This personalization would add a uniquely intimate layer to the interaction, similar to how secret handshakes signify closeness between two people or some in-group status. 

\subsubsection{Accessibility limitations.}

\textcolor{black}{Currently, \sys is limited in accommodating all abilities (e.g., people with mobility impairments and those with blindness or low vision). Feminism should acknowledge intersectionality, and that includes building for various abilities. Expanding the range of gesture-based methods to accommodate users with different abilities, needs, and various device models is crucial. For individuals with limited hand dexterity or those who may not be able to perform certain gestures, providing customizable gestures or alternative input methods, such as voice commands or facial expressions, could enhance usability. For users with visual impairments, integrating haptic feedback or audio cues to guide and confirm interactions can make the system more accessible. Compatibility with screen readers and providing clear, descriptive labels and instructions for all gestures and functions would also be important. Accessibility testing with diverse user groups may be a continuous part of the development process to identify and address any barriers that may arise. By prioritizing inclusive design, \sys can better serve a broader range of users and embody principles of intersectional feminism.}

\section{Study}
Following our design goals outlined in Section \ref{designgoals}, we set out to answer these questions:

\begin{enumerate}
    \item[\textbf{RQ1.}] Can hand gestures be completed \& recognized successfully? 
    \item[\textbf{RQ2.}] Can hand gestures deter screenshots? 
    \item[\textbf{RQ3.}] What are peoples' preferences for hand gestures under different situations?
\end{enumerate}

RQ1 draws from \textbf{D1} to evaluate the effectiveness of the gesture recognition model and whether participants are able to perform the gestures. RQ2 speaks to \textbf{D2} to focus on the deterrence ability. RQ3 assesses participant preferences for \textbf{D3}. RQs 1 and 2 are evaluated via in-lab usability tasks, and we answer RQ3 through interviews with the same participants. This research was reviewed by the University of Michigan Institutional Review Board and received an exemption as a benign behavioral intervention.

\subsection{Materials and procedure}
We designed an experiment using two tasks to answer our research questions. Participants first received a 3-minute demonstration of \sys to ensure that they were physically able to perform each gesture.  The first task focused on participants' ability to execute the gestures. The second task focused on participants' ability to circumvent the \sys system. 

For the first task, an iPhone running the \sys iOS app was placed in front of the participant. The gestures were shown in random order and participants had up to one minute to perform each gesture to have it recognized by the model \textit{once}. After a gesture is detected, \sys will move on to the next gesture. We recorded videos of participants while using \sys and documented for each gesture the seconds that it took for the gesture to be detected by the system. 

The second task asked the participant to try to take a screenshot of the unlocked screen to the best of their ability. We instructed them to take screenshots by pressing two buttons on the iPhone. Naturally, it is difficult to hold a hand gesture and also take a screenshot at the same time, so we instructed the participants to use whatever method they can to ``break'' the system and try to take the photo. The participants received no additional hints or instructions beyond this. For each gesture, participants are given up to 3 minutes or until they choose to give up and move on to the next gesture. We again recorded videos of the participants, and for each gesture, we documented the number of attempts. 

\vspace{5pt}
Participant results for the screenshot deterrence task can be broken up into the following categories:

\begin{itemize}
    \item \textbf{Successful} screenshot capturing an unlocked screen.
    \item \textbf{Failed} can be defined in two ways:
        \subitem \textbf{Attempted} screenshot but the screen became locked due to the gesture not being detected when a screenshot was taken.
        \subitem \textbf{Skipped} the trial without making an attempt; participant believed it was ''impossible" to take the screenshot. 
\end{itemize}

After each task, we asked participants to complete the NASA Task Load Index (NASA-TLX) to compare the difficulty between the two tasks. NASA-TLX is a long-standing measurement of mental workload with 21 questions on the mental and physical demands of the activity, temporal demand, performance, effort, and frustration experienced in the task~\cite{hart1988development, hart2006nasa}.

At the end of the evaluation, we asked each participant the following open-ended interview questions to assess how well gestures satisfy the varying suitability guideline (\textbf{D3}):
\begin{itemize}
    \item Imagine you were messaging a friend and they sent a message that you had to unlock with a hand gesture. Would you assume they would be okay for you to screenshot it?
    \item Which hand gesture are you most likely to use to send: 
    \subitem \dots something funny to a close friend?
    \subitem \dots something personal you don't want to be opened in public?
    \subitem \dots something serious?
    \subitem \dots something to someone you're not close with? 
\end{itemize}

The evaluation was performed in a private, mostly empty office set up for the experiment design. The participants sat at a desk and used a provided iPhone 15 (iOS 17.6) on which \sys was installed. A stand attached to the case supports the phone, but the phone can also be laid flat on a table. We recorded video in two different ways: (a) a screen recording captured the exact on-screen interaction between the participant and \sys; and, (b) a camera on a separate laptop captured the broader physical interaction of the participant and the device.

\subsection{Participants}
We invited 13 participants to attempt to take screenshots while using \sys. We selected participants who currently own and use an iPhone as their main mobile device. The participants consisted of undergraduate and graduate students whose ages ranged from 18 to 30 years old ($\mu$ $=$ 23.8 years). We did not specifically restrict participation based on physical abilities or limitations (though individuals may have self-selected to participate).  All 13 were able to successfully perform the hand gestures without assistance. Each participant was compensated \$20 following the study.

\subsection{Results}
\subsubsection{Gestures as deterrence}
\textcolor{white}{.}\vspace{0.1cm}

\begin{figure}[t]
    \centering
    \includegraphics[width=0.95\linewidth]{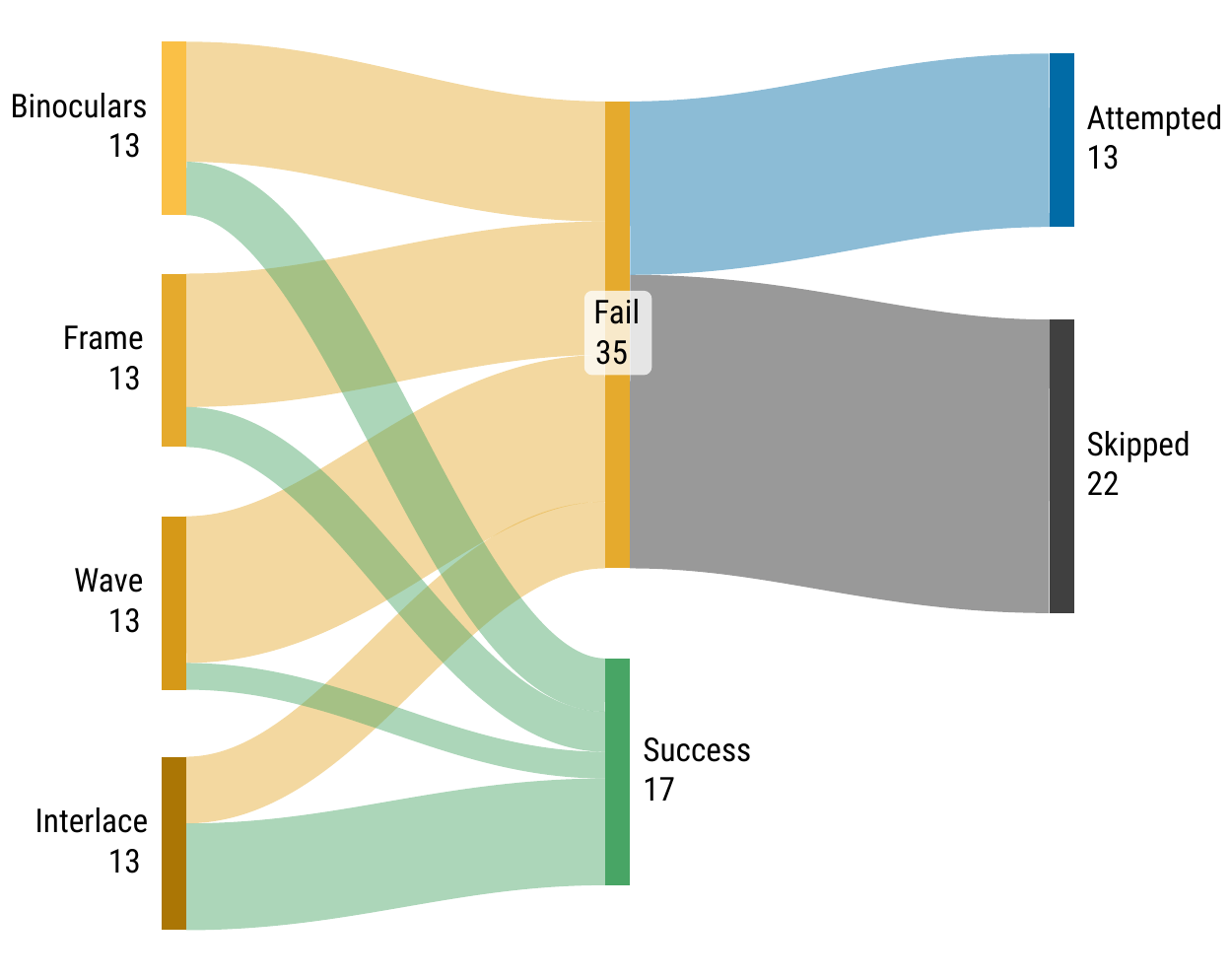}
    \captionof{figure}{Of all attempts, 33\% succeeded in making a screenshot, 25\% attempted without success, 42\% gave up early on believing the task to be impossible.}
    \label{fig:sankey}
\end{figure}

Using \sys deterred 67\% of all screenshot attempts. The success rate increases to 77\% when we take out the interlace gesture. Figure \ref{fig:sankey} shows the results across all usability tasks. On average, participants took between 1 and 11 seconds ($\mu$ $=$ 2.78 sec) to perform a hand gesture and to have it detected by the system. In contrast, when attempting to screenshot while performing a hand gesture, the mean time spent across all hand gestures was 74.4 seconds and up to the full time allotment of 180 seconds. Notably, in 7 trials participants chose to skip a gesture when they felt it was impossible to ``break'' and decided to give up and move on to the next gesture without making an attempt. 

\begin{figure}
    \centering
    \includegraphics[width=0.8\linewidth, clip, trim=15pt 10pt 10pt 15pt]{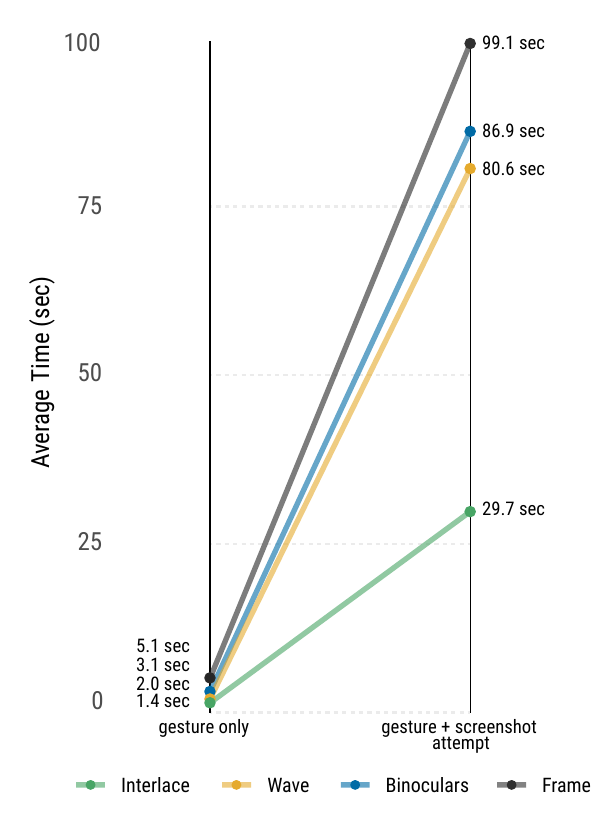}
    \captionof{figure}{Completion times for the two evaluation tasks. Performing gestures was easy (left), but trying to screenshot while performing gestures was difficult (right).}
    \label{fig:slope}
\end{figure}

\begin{figure*}
    \centering
    \includegraphics[width=0.85\linewidth, clip, trim= 2pt 0pt 6pt 18pt]{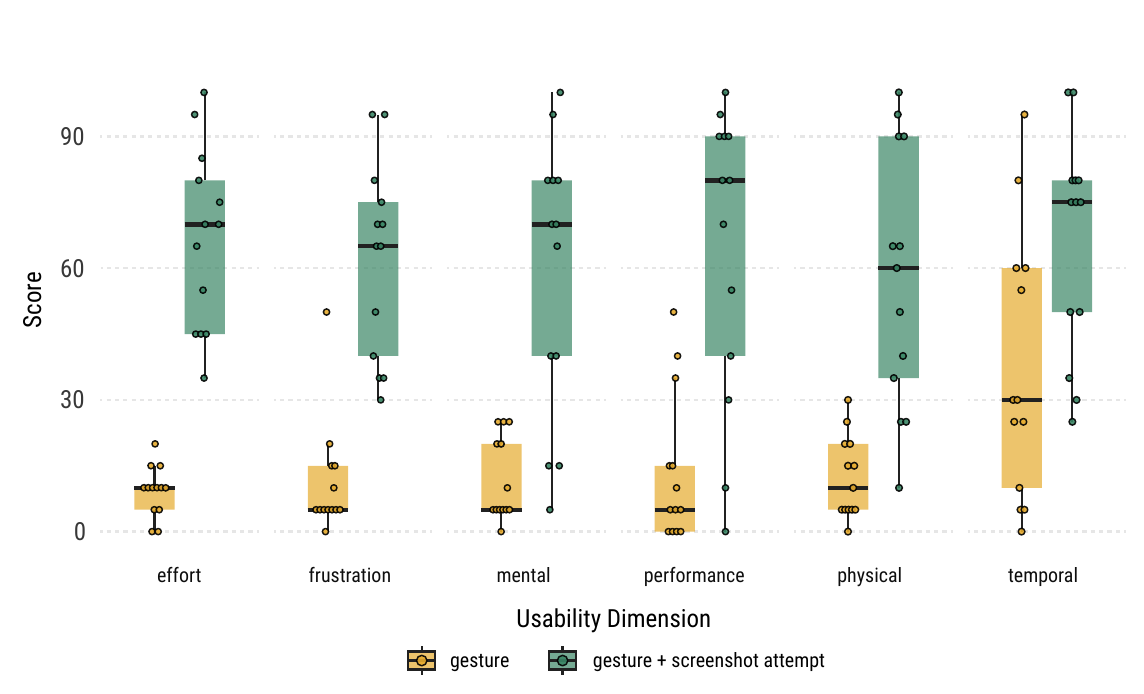}
    \captionof{figure}{Participants' self-reported NASA-TLX usability results show it is easy to perform \sys gestures on their own, but difficult do gestures and attempt to screenshot at the same time.}
    \label{fig:usability}
\end{figure*}

Participants were able to perform the hand gestures on their own easily, but found it very difficult to take a screenshot. Even when we exclude participants who chose to give up early, Figure \ref{fig:slope} displays the dramatic difference between time spent on the first and second tasks. The time spent screenshotting was, on average, 27 times longer than that doing hand gestures only. The majority of successful screenshots were using interlace due to the aforementioned loophole. Similarly, the self-reported usability metrics in Figure \ref{fig:usability} suggest that participants found that performing the gestures was easy, but that attempting screenshots was difficult.

\subsubsection{Getting around \sys} 
\textcolor{white}{.}\vspace{0.1cm}

\textbf{Playing with the model.} Participants prodded, poked, and pushed the gesture recognition model to find its working boundaries. Participants attempted different angles and camera orientations to find the limits of the model. One participant creatively attempted to ``trick'' the model with his index finger joints posing as a frame (without success). We also witnessed attempts to make the gesture to unlock the phone and then quickly try to screenshot. We found that the interlace gesture was a special case. With the exception of interlace, no other attempts to push the boundaries of the model resulted in successful screenshots. 

\textbf{Using parts of the body.} With the hands engaged in performing the gesture, some participants tested alternative parts of the body to press the two side buttons on the iPhone. This included elbows, knees, and even feet. The use of feet and knees was successful because the hands could be free to perform the gesture. Some gestures, such as interlace and binoculars, did not engage all fingers (e.g., free pinkies or thumbs), which one participant was able to use to screenshot. 

\textbf{Using technologies.} For the purposes of the usability testing, we did not disable voice interactions (i.e., Siri). A few savvy participants used this to take a screenshot, resulting in some of the fastest successful times. We note that this feature could be blocked in the application.

\textbf{Using other objects.} The usability laboratory contained a few quotidian items, including snack bars and small empty cardboard boxes. Participants were told they could use any method they could think of to attempt to screenshot. Using a box to press the button was unsuccessful, as was usage of the snack bars.

\vspace{1cm}

\subsubsection{Gestures as social signals} 
\textcolor{white}{.}\vspace{0.1cm}

\textbf{Sent with gestures == don't screenshot me!} Nearly all participants noted the social significance of a message sent with a hand gesture. The additional friction of performing hand gestures pushed participants to consider the context in which the exchange was taking place. In their statement, P13 considered the sender's desired permanence and understood the use of gestures to be ``\textit{something similar to the concept of invisible ink. If it's something you have to do another task to see, I'm assuming that it's something super private, and it's probably not meant to be kept for a long time.''}. While taking a screenshot is not entirely impossible, the difficulty of the task is effective for starting a discussion about the appropriateness of taking a screenshot. P1 says, ``\textit{there's a social presumption of safety and privacy. To take a screenshot is to go against the social mores of the technology. It feels bad to do it, so any kind of buffering you can do to force people into a social more is a good idea. It forces you to think about something before you do it.''}

\begin{figure}[t]
    \centering
    \includegraphics[width=0.95\linewidth]{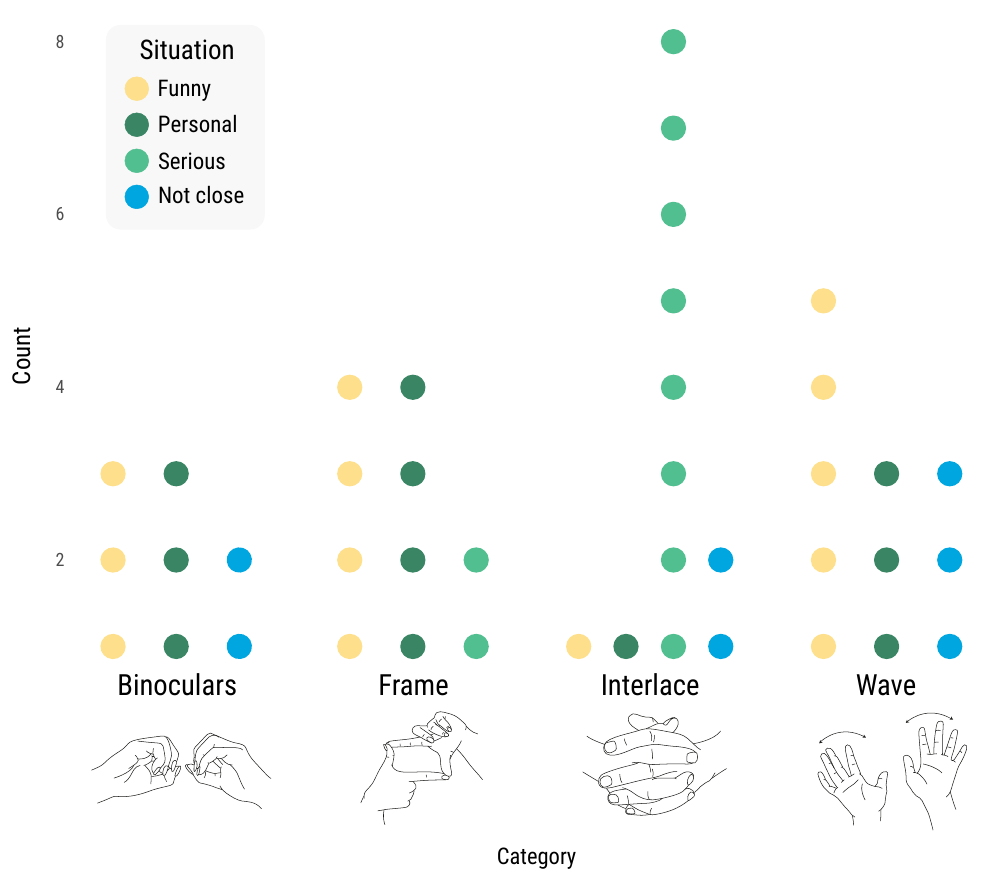}
    \captionof{figure}{Strong preferences for using interlace in ''serious" situations and mixed preferences for other situations.}
    \label{fig:situations}
\end{figure}

\textbf{Hand gestures as an emotional language.} We asked participants to select the gesture they believed to be most appropriate under different social contexts (see Fig. \ref{fig:situations}). Responses showed clear patterns to the appropriateness of gestures in different situations. 

\noindent We note some flexibility in interpretations because there are often multiple interpretations of the emotional value of a gesture. 

One approach to choosing the best gesture for sending a funny photo to a close friend was to take into account the context in which the photo is being sent, the sender, and the media. P12 chose the frame gesture because it mirrored ``\textit{taking a picture, and you get to see my funny photos when using the picture.''} A different approach more heavily considered which gesture would be the simplest for the recipient to complete, considering that a funny photo to a close friend calls for a low-security measure. Due to the underlying hand recognition loophole that allowed \textit{interlace} to be screenshotable with one hand, P10 believed interlace ``\textit{was easy enough to only have to use one hand to do it. That's the best preferred, only because you can trick the system with that one.''}

Interlace and frame were top choices for gestures when participants would not want the recipient to open the message in public, but for possibly different reasons. The physical gestures of interlace implied seriousness. P1 supports this by saying that \textit{``interlaced fingers would be the one that I would pick, just because it is stern and it's closed...It's almost like you're holding something, cradling something that you wouldn't want seen.''} In contrast, P11 selected frame to deter the recipient from opening the message in public spaces: \textit{``If it's in public, I would want it to be a gesture that would be weird to do in public, so I feel the frame is pretty strange with doing it to your phone. You might wave at your phone or you might set up your phone on a little stand or something and lace your fingers in front of it. Both those things seem more natural, so they're easier to do in public.''} 

For serious content, some participants preferred interlace. P13 believes \textit{``because interlace is the easiest one to maintain while also still staying focused on what you're looking at.''} Participant 1 noted that interlace had a serious nature to it: \textit{``This is what I would do with a job interview, at something serious like a stern meeting. If I'm arrested, I put my hands like this.''}

Although most of the participants had clear gesture preferences for defined situations, others such as P2 believed all gestures were fine for any situation: \textit{``All of these gestures don't seem overly silly, where it'd be weird if you had to make a weird face or something, but since it's just a gesture to open a message, I feel like they're all appropriate in that way.''}

\subsection{Summary}

To recap, we summarize the answers to our evaluation questions. 

\begin{itemize}
    \item[\textbf{RQ1.}] \textbf{Can the hand gestures be completed and recognized successfully?} Yes, on average it took fewer than 3 seconds for a model to recognize a hand gesture. See Figure \ref{fig:slope}.
    \item[\textbf{RQ2.}] \textbf{Can hand gestures deter screenshots?} Two-thirds, or 67\% of screenshot attempts, resulted in failure. Restricting to the top-performing three gestures (e.g., eliminating interlace), the deterrence rate was 77\%. Among successful attempts, we see interesting (and socially unlikely) methods of trying to screenshot. See Figures \ref{fig:sankey} and \ref{fig:task2}.  
    \item[\textbf{RQ3.}] \textbf{What are peoples' preferences for hand gestures under different situations?} Many considered the emotions the gestures evoked and the practicality of using certain gestures in different situations. For example, interlace is largely preferred in serious situations. See Figure \ref{fig:situations}.
\end{itemize}

\section{Discussion}
\sys reduced non-consensual screenshots, and represents a type of feminist interaction technique. To conclude, we attempt a conceptual contribution, and move toward generalizing the approach of \sys. We discuss how interaction techniques can be feminist, and the opportunities and constraints associated with such an aim. 

\subsection{Feminist Interaction Techniques}
\begin{figure}[t]
    \centering
    \includegraphics[width=\linewidth , clip, trim= 2pt 230pt 75pt 18pt]{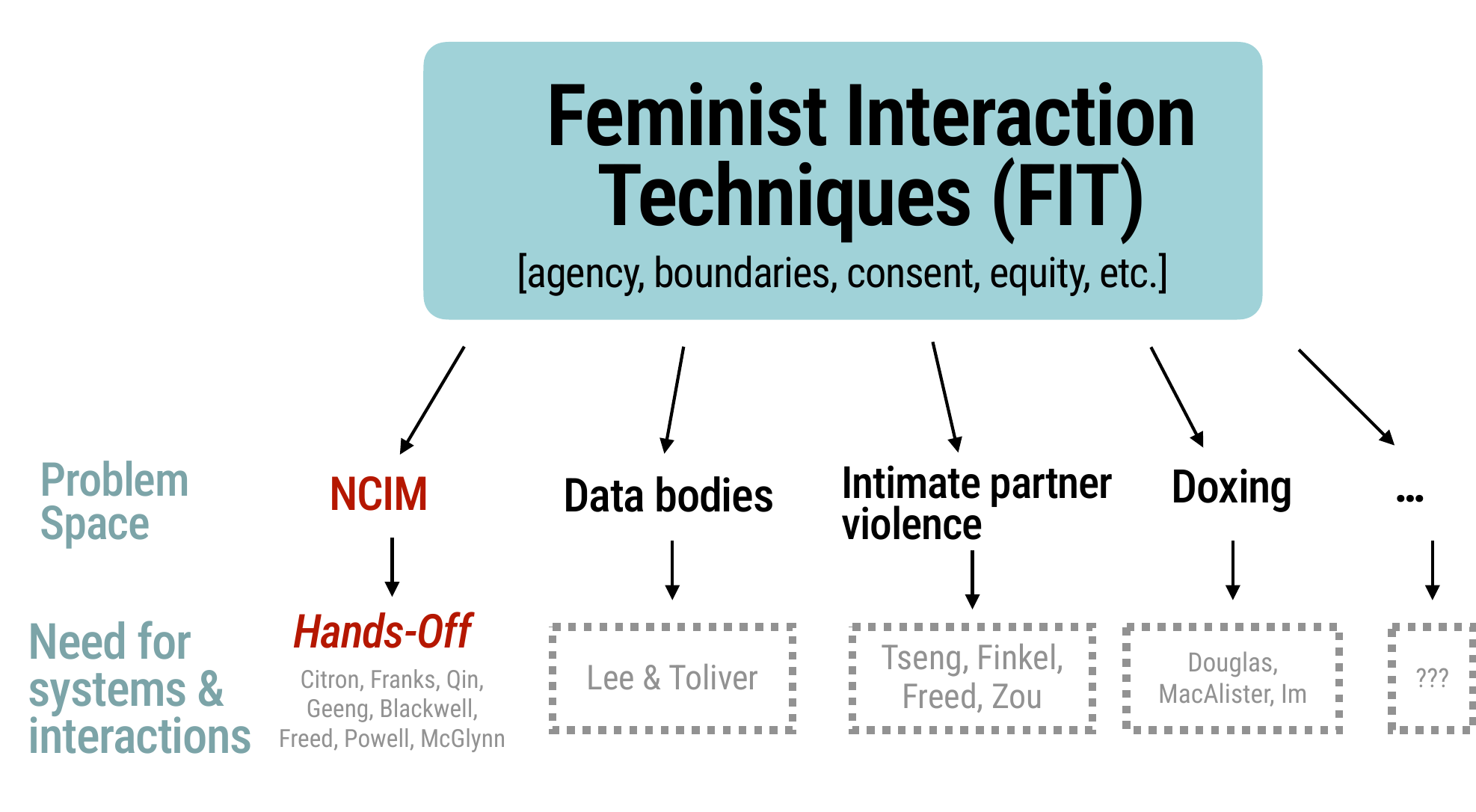}
    \captionof{figure}{A conceptual framework of FIT, incorporating other scholars who have called for systems and interaction techniques in domains such as data bodies (e.g,~\cite{lee_building_2017}), intimate partner violence (e.g.,~\cite{freed2018stalker, tseng2020tools, zou2021role}), and doxing (e.g.,~\cite{im_yes_2021}).}
    \label{fig:fit}
\end{figure}

This paper introduces the concept of \textit{Feminist Interaction Techniques (FIT)}---interaction techniques that embody feminist values and speak to societal problems (see Fig. \ref{fig:fit}). While some subdomains of computing (e.g. the field of Social Computing \cite{steinhardt2015feminism}) have been receptive to the idea of feminism-inspired designs, systems built with such values are few and far between. Part of the problem is the inherent difficulty in mapping high-level concepts to concrete system interactions. With FIT, we aim to bridge the gap between what values systems should have and how they can be built into interactive technologies. We address calls to make feminism concrete in systems:``\dots while feminism is generally known as a critical strategy, which too often suggests that feminism is only applicable after the fact \dots [we] stress the potential for feminism to contribute to an action-based design agenda''~\cite{bardzell_feminist_2010}.

Our application of FIT to the NCIM space demonstrates a way to overcome the limitations of existing solutions. Current NCIM solutions rely primarily on reporting violations; however, reporting is burdensome for users. It requires interacting with an unknown third party (whoever reviews the report for a platform), and can even require submitting \textit{more} photos of oneself and photos of IDs for identification or matching purposes \cite{geeng_usable_nodate,mcglynn_its_2021,de_angeli_reporting_2023}. These approaches are `state-of-the-art' in deployed systems, but they are not particularly feminist. The \sys system puts the power to establish boundaries in the hands of the photo owner rather than the recipient. It also shifts how we think about \textit{who} should bear the burden of consent from the victim to the potential violator. Finally, it adopts a proactive, rather than reactive, approach, allowing people to establish boundaries rather than reacting to harms.

\subsection{FIT values}
Feminism is a multi-faceted movement and discipline that has evolved through different waves and phases (see a review within the computing context in~\cite{bardzell_feminist_2010}). However, many of its core values are durable, including consent, boundaries, equity, bodily autonomy, economic justice, and racial justice. In the case of NCIM, an overarching value is first that NCIM is a problem worth addressing--it affects millions or billions of people daily and degrades social life online and offline. In focusing on NCIM, we also embrace the values that people should have control over their likeness online. Sharing content of someone's likeness without their consent can be a violation of their bodily rights and autonomy, and the field of computing bears some responsibility for addressing the problem of non-consensual sharing of content.

While outside the scope of this paper to exhaustively list values, FIT can leverage feminist values to create new interaction techniques. Critically, our user study suggests that we can effectively design systems that communicate these values to users. As one participant said after using \sys, ``There's a social presumption of safety and privacy \dots It forces you to think about something before you do it.'' We could also co-create values with users. In this study, we pre-selected the four hand gestures for participants, which is appropriate for early exploration~\cite{wobbrock2009user}. However, as shown in \ref{fig:designspace}, the hand gestures bore social meaning, e.g. a wave is more socially acceptable than a binoculars gesture. 

Future work could run gesture elicitation studies~\cite{villarreal2024brave}, allowing people to create gestures aligned with their own values. There could also be libraries of consent-based gestures by aggregating preferences across many users \cite{morris2010understanding}

\subsection{Application spaces for FIT}
Feminist interaction techniques have many use cases. Anywhere there are societal problems that introduce threats to safety and well-being, we can bring a feminist-inspired lens to the problem. The decision to do so does not introduce bias to the design space, but instead recognizes that many social problems are themselves rooted in bias. The power of feminist interaction techniques is that it does not take away from one group in order to design for another; a rising tide lifts all boats.

Digital safety, and other problem spaces with interpersonal and social harms, are natural fits for FIT; however, FIT is not limited to these contexts. There has been much demand for systems and mechanisms that make online interactions safer, more inclusive, and more participatory. It is possible to think of many different vectors for applying FIT to common social interaction paradigms. How to do this in different domains remains an open question---and an opportunity for creative technical HCI research.

A generative approach may be to apply a feminist value to an interaction technique of interest and ask, ``How are assumptions about this value `baked into' it?" In the case of \sys, we targeted the consent assumptions built into the idea that once an image is sent to someone, they have immediate and irrevocable access to it. How are concepts of bodily autonomy (or lack thereof) baked into the context of meeting someone's VR avatar? How is racial justice encoded in the techniques we use to explore spaces in AR---given that the history of many locations and spaces are contested (e.g., Indigenous lands)? What assumptions about consent does deepfake technology make?

\subsection{Limitations and precautions}

The computing community has a role to play and a responsibility to bear to address digital harms. Problems like NCIM would not exist without decades of computing work that put cameras and mobile devices into nearly everyone's hands, with few safeguards on their use. However, it is important to interrogate the limits of FIT---and any technical approach---in important societal problems. FIT cannot solve gender-based violence. No technology can. Problems like NCIM are technologically-mediated, but the social and societal contexts in which they exist have been around for millennia. The oppression of women and gender minorities, how the law treats these harms, how society treats survivor-victims, etc., are problems that technology cannot single-handedly ``fix.'' For computing communities with techno-solutionist norms, it is important to note that no technology---no matter how clever and well-intentioned---will completely solve complex societal problems. Additionally, feminism itself, like any set of values or principles, has flaws and limitations (e.g., early feminism prioritized gender while overlooking race, a pattern that was also observed with social technologies \cite{steele2021digital}).

We began the paper with the question: can interaction techniques be feminist? Our \sys system demonstrates how they can be, but we should proceed with caution in this goal. System design seeks to abstract and concretize outcomes, but consent is a practice not an outcome. To realize the goal of feminist values in interaction techniques, we should not overfit to a design space of solving an interaction problem, but instead to the goal of augmenting an interaction process aligned with desired values. 

\begin{acks}
    This material is based upon work supported by the National Science Foundation under Grants 1763297 and 2311102.
\end{acks}

\balance
\bibliographystyle{ACM-Reference-Format}
\bibliography{sample-base}

\end{document}